\documentclass[twocolumn,showpacs,preprintnumbers,amsmath,amssymb]{revtex4}
\usepackage{graphicx}
\usepackage{dcolumn}
\usepackage{bm}

\begin{document}
\preprint{...}
\title{Mechanical stiffening, bistability, and bit operations in a microcantilever}
\author{Warner J. Venstra} \email{w.j.venstra@tudelft.nl}
\author{Hidde J.R. Westra} 
\author{Herre S.J. van der Zant}
\affiliation{Kavli Institute of Nanoscience, Delft University of Technology, Lorentzweg 1, 2628CJ Delft, The Netherlands}%
\date{\today}
\begin{abstract}
We investigate the nonlinear dynamics of microcantilevers. We
demonstrate mechanical stiffening of the frequency response at large
amplitudes, originating from the geometric nonlinearity. At strong
driving the cantilever amplitude is bistable. We map the bistable
regime as a function of drive frequency and amplitude, and suggest
several applications for the bistable microcantilever, of which a
mechanical memory is demonstrated.
\end{abstract}

\maketitle

\indent\indent Microcantilevers are widely applied as transducers in
sensitive instrumentation \cite{Rugar04,Harris09}, with scanning
probe microscopy as a clear example. Typically, the cantilever is
operated in the linear regime, i.e. it is driven by a harmonic force
at moderate strength, and its response is modulated by the parameter
to be measured. In clamped-clamped mechanical resonators, additional
applications have been proposed based on nonlinear behavior.
Nonlinearity in clamped-clamped resonators is due to the extension
of the beam, which results in frequency pulling and bistability at
strong driving, and can be described by a Duffing equation
\cite{Kozinsky06}. Applications which employ this bistability are
e.g. elementary mechanical computing functions \cite{Badzey04,
Mahboob08}. Since a cantilever beam is clamped only at one side, it
can have a nonzero displacement without extending. One would
therefore not expect a Duffing-like behavior for a cantilever beam.
Nonlinear effects of a different origin have been observed in
scanning probe microscopy, due to interactions between the
cantilever and its environment. Tip-sample interactions either
weaken or stiffen the cantilever response, depending on the strength
of the softening Van der Waals forces and electrostatic interactions
and the hardening short range interactions
\cite{Ruetzel03,Mueller99}. Weakening also occurs when the
cantilever is driven by an electrostatic force \cite{Kacem10}.
Besides nonlinear interactions with the environment, theoretical
studies predict \emph{intrinsic} nonlinear behavior of cantilever
beams \cite{Crespo78a,Crespo78b, NayfehMook, Mahmoodi07, Kacem10}
, of which indications have been reported \cite{Mahmoodi07,Ono08}.\\
\indent\indent In this letter, we report a detailed experimental
analysis on the nonlinear mechanics of microcantilevers. It is shown
that a hardening geometric nonlinearity dominates over softening
nonlinear inertia, which effectively leads to a stiffening frequency
response for the fundamental mode. At large amplitudes, the
mechanical stiffening results in frequency pulling and ultimately in
intrinsic bistability of the cantilever. We study the bistability in
detail by measuring the cantilever response as a function of the
frequency and amplitude, and compare the experimental observations
with theory. A good agreement is found. We suggest several
applications for the bistable cantilever, and as an example we
demonstrate that bit operations can
be implemented in the bistable cantilever.\\
\begin{figure}[b]
\includegraphics[width=85mm]{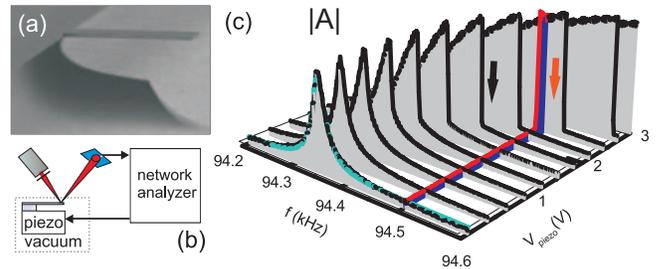}
\caption{(a) Scanning electron micrograph of a silicon nitride
cantilever; (b) Experimental setup; (c) Response lines for several
drive voltages (forward frequency sweeps). A damped driven harmonic
oscillator fit is shown for the weakly driven cantilever. The line
at $\mathrm{f=94.5 kHz}$ represents a response along the
(decreasing) drive strength axis. The arrows indicate the switching
direction.}
\end{figure}
\indent \indent Experiments are performed on thin cantilevers with a
rectangular cross section, $\mathrm{w \times h}$, fabricated from
low-pressure chemical vapor deposited silicon nitride using electron
beam lithography and an isotropic reactive ion etching release
process. Figure 1 (a) shows a scanning electron micrograph of a
fabricated cantilever. The cantilever is mounted on a piezo actuator
and placed in a vacuum chamber at a pressure of $\sim10^{-4}$ mbar.
At this pressure, the cantilever operates in the intrinsic damping
regime. An optical deflection technique is deployed to detect the
displacement of the driven cantilever, and the frequency response is
measured using a network analyzer, see Fig. 1(b).\\
\indent\indent Figure 1 (c) shows frequency response lines for a
weakly and strongly driven cantilever with length $\mathrm{L=40\;\mu
m}$ and $ \mathrm{w \times h = 8\;\mu m \times 200\;nm}$. For weak
driving the response fits a damped driven harmonic oscillator, with
$\mathrm{f_0=94.35}$ kHz and $\mathrm{Q \approx 3000}$. Figure 1(c)
also shows the response when driven at increasing strength: the
resonance peak shifts to a higher value and the response becomes
bistable. It resembles the response of a clamped-clamped beam driven
in the nonlinear regime. A more detailed measurement is presented in
Fig. 2(a,b). Here the magnitude of the resonator response, $|A|$, is
depicted (color scale) as a function of the drive frequency and
strength. The frequency is swept forward (i.e. from a low to a high
frequency, FW) and backward (BW), and after each frequency response
measurement the drive strength is increased. Parameters which result
in a hysteretic (HY) response are visualized by subtracting
forward and backward traces, as shown in Fig. 2(c).\\
\begin{figure}[t]
\includegraphics[width=85mm]{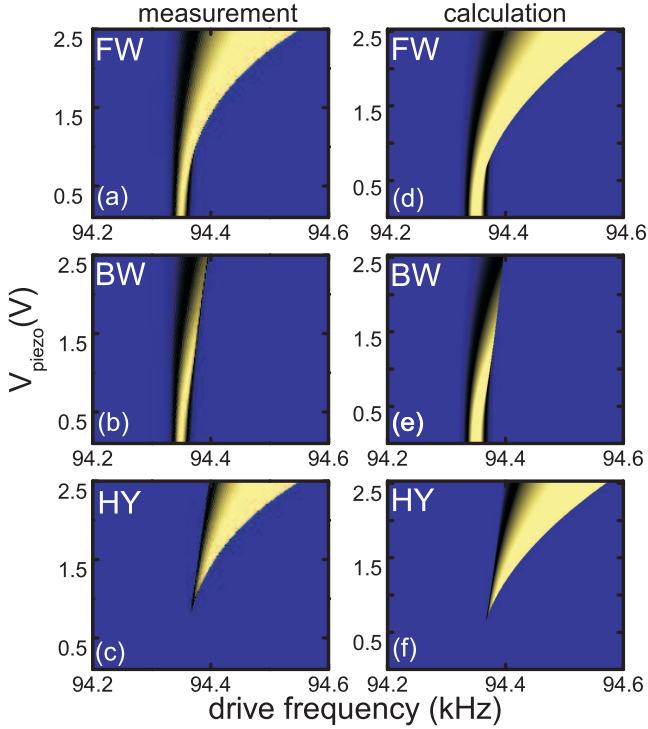}
\caption{FIG. 2. Frequency pulling and bistability in a cantilever,
measurement (left) and calculation by solving Eq. 3 (right). The
color scale represents the magnitude of the frequency response,
$\mathrm{|A|}$. The drive frequency is swept from a low to a high
value (a,d) and vice versa (b,e). Panels (c,f) show the bistable
regime, obtained by subtracting the forward from the backward
response. As the piezoelectric coupling parameter is not known, the
y-axis in the calculations has been scaled to match the experimental
values.}
\end{figure}
\indent\indent The theory of nonlinear oscillations of a cantilever
beam has been developed in Refs. \cite{Crespo78a,Crespo78b}. Using
the extended Hamilton principle the equation of motion for the
displacement $\tilde{u}$ is derived:
\begin{widetext}
\begin{eqnarray}
D [\tilde{u}'''' + [\tilde{u}'(\tilde{u}'\tilde{u}'')']'] + \rho w h
\ddot{\tilde{u}} + \tilde{\eta} \dot{\tilde{u}} + \frac{1}{2} \rho w
h \Big( \tilde{u}'\int_L^s \frac{\partial^2}{\partial \tilde{t}^2}
\int_0^{s_1} (\tilde{u}')^2 ds_2 ds_1 \Big)' = \tilde{F}.
\end{eqnarray}
\end{widetext}
The dots and primes denote differentiation to time $\tilde{t}$ and
the arc length $s$ of the cantilever respectively, and $D$ is the
bending rigidity, $\rho$ the density, and $\tilde{\eta}$ is the
damping parameter. The piezo actuator generates a displacement
${U}=d_{33} V \cos(\tilde{\Omega} t)$, where V is the drive voltage
and $d_{33}$ the piezoelectric coefficient. The resulting force on
the cantilever equals $\tilde{F}=\ddot{U}\rho w
h=-\tilde{\Omega}^2\rho w h d_{33} V \cos(\tilde{\Omega} t)$.
Equation (1) is transformed to a dimensionless form by substituting
$u = \tilde{u}/h$, $x = s/L$, $l = d_{33}V/h$, $\eta = \eta'L^4/(D
\tau)$ and $\delta = (h/L)^2$. The time $\tilde{t}$ and drive
frequency $\tilde{\Omega}$ are scaled using $\tau = L^2 \sqrt{\rho w
h/D}$. Applying the Galerkin procedure \cite{Atluri73, Kacem10} for
the first mode ($u = a(t)\xi(x)$) gives:
\begin{equation}
\ddot{a} + \omega^2 a+ \eta \dot{a} + 40.44 \delta a^3 + 4.60 \delta
(a \dot{a}^2 + a^2\ddot{a}) = - 0.78 l\Omega^2 \cos(\Omega t).
\label{eq:dimless}
\end{equation}
Here, $a$ is the normalized coordinate, and $\omega$ the
dimensionless resonance frequency; for the first mode $\omega =
3.52$. The cubic term in $a$ represents the hardening geometric
nonlinearity, and the fifth term represents nonlinear inertia which
softens the frequency response \cite{Kacem10}. The values 40.44,
4.60 and 0.78 are obtained by integrating the linear mode shapes,
$\xi(x)$~\cite{integrals}. Equation~\ref{eq:dimless} can be solved
using the method of averaging or the method of multiple scales
\cite{NayfehMook} and the amplitude, $A$, can be implicitly written
as:
\begin{equation}
A = \frac{l \Omega^2}{\sqrt{6.57(15.16 \delta A^2 - \omega \Omega + \omega^2(1 - 1.15 \delta A^2))^2 + 1.64 \eta^2 \omega^2}}.
\end{equation}
This equation is solved self-consistently to obtain the resonator
amplitude, which is normalized by the drive strength $l$ to obtain
the frequency response. Using the experimentally obtained linear
resonance frequency, Q-factor and the dimensions as input
parameters, the frequency responses are calculated as a function of
the drive strength. Figures 2 (d,e) show the simulated stable
solutions, which correspond to the resonator response to a forward
and backward frequency sweep. The model captures the observed
behavior well, where the piezoelectric coupling parameter is the
only free parameter. Both the calculations and the experiments
indicate that the geometric nonlinearity dominates over the inertial
nonlinearity. Analyzing Eq. (3) in detail shows that the
nonlinearity depends on the modeshape, $\mathrm{\xi(x)}$, and the
squared aspect ratio, $\delta$. For the fundamental mode, the
intrinsic nonlinearity in cantilevers always leads to stiffening of
the frequency response. In contrast, the calculation shows that the
same nonlinearity results in a weakening effect for higher
modes \cite{highermodes}.\\
\indent\indent The intrinsic mechanical bistability allows
cantilever applications similar to the ones implemented in
clamped-clamped resonators. As an example, we demonstrate mechanical
bit operations in a cantilever with dimensions $L \times w \times h
=\,\mathrm{ 30\,\mu m \, \times\,8 \, \mu m\, \times\,150}$ nm, with
a linear resonance frequency $\mathrm{f_0=193.49}$ kHz and
$\mathrm{Q \approx 5800}$ in vacuum. For this cantilever, a
measurement of the hysteretic regime is shown in Fig 3 (a)
\cite{piezo}. Bit operations can be performed by modulating the
drive frequency or the drive strength --or a combination thereof--
across the hysteretic regime, a scheme that was also deployed to
implement nanomechanical memory in clamped-clamped beams
\cite{Badzey04, Mahboob08}. The principle is indicated by the arrows
in Fig. 3(a). The drive strength is modulated by varying the voltage
on the piezo at a fixed frequency, as shown in Fig. 3(b). A backward
sweep in the drive strength follows the high-amplitude state,
similar to a forward sweep in drive frequency. This intuitively
becomes clear in Fig. 1(c), where the transition from a high to a
low amplitude occurs during a backward sweep in the drive strength,
as indicated by the red line along the fixed frequency at
$\mathrm{f=94.5 kHz}$. During a forward sweep in the drive strength
the resonator follows the low-amplitude stable branch, as with a
backward sweep in
frequency.\\
\begin{figure}[t]
\includegraphics[width=85mm]{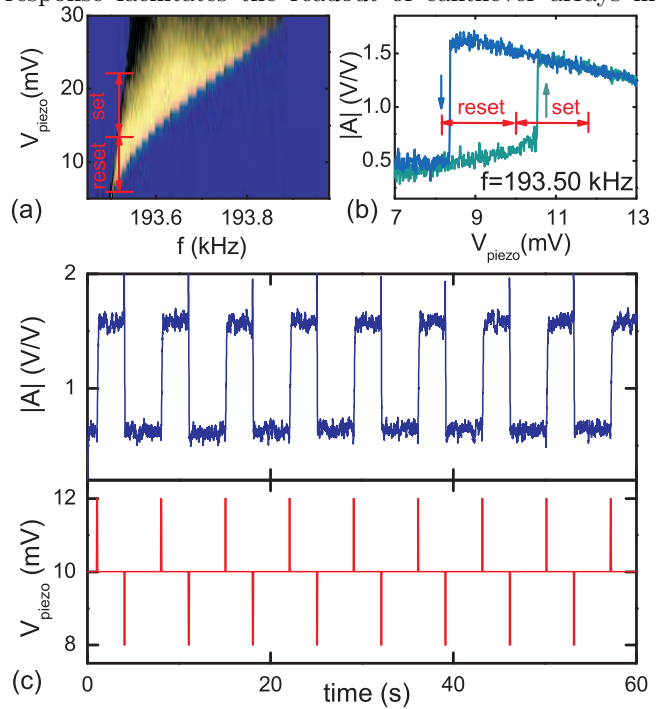}
\caption{Hysteretic regime for a $\mathrm{30\;\mu m \times 8\;\mu m
\times 150\;nm}$ cantilever. (b) Drive strength sweep at fixed
frequency, and indication of the modulation to implement the bit.
(c) Mechanical memory in a bistable cantilever beam: drive strength
(lower panel) and cantilever response (upper panel).}
\end{figure}

\indent\indent To implement the bit, the cantilever is driven in the
bistable regime at $\mathrm{f=193.50\;kHz}$ and $\mathrm{V_{piezo} =
10\;mV}$.  To set and reset the cantilever bit, the drive voltage is
modulated by 2 mV around the operating point, as indicated by the
arrows in Fig. 3(b). Starting at low amplitude, '0' in Fig. 3(c), a
high-amplitude '1' is written by temporary increasing the drive
voltage to 12 mV. The cantilever switches to a high vibrational
amplitude and remains in this state after the drive voltage is set
back to the operating point. Next, the drive strength is lowered to
8 mV which resets the cantilever to a low amplitude oscillation,
corresponding to '0'.\\
\indent\indent Bistability of cantilever beams can be used for
various purposes besides the mechanical memory application described
here. For example, the hysteretic frequency response facilitates the
readout of cantilever arrays in dissipative environments  by
employing the scheme described earlier \cite{Venstra08}. Bistability
may also open the way to use a cantilever as its own bifurcation
amplifier \cite{Greywall94,Siddiqi04, Gammaitoni98,Almog07} in for
example scanning probe microscopy, thereby enhancing the sensitivity
to external stimuli. Finally, we note that despite scaling with the
aspect ratio squared, $\delta$, the bistable regime is also
accessible for single-clamped nanoscale resonators such as carbon
nanotubes \cite{Perisanu10}.\\
\indent\indent In conclusion, we investigated the nonlinear
oscillations of microcantilever beams. Mechanical stiffening is
observed which results in frequency pulling and bistability. The
experiments are in excellent agreement with calculated nonlinear
response. Several applications for the bistable cantilever are
suggested, of which a mechanical memory is demonstrated.\\
\indent\indent The authors acknowledge Khashayar Babaei Gavan for
fabricating the devices, and the Dutch organization FOM (Program 10,
Physics for Technology) for financial support.

\begin{thebibliography}{23}
\bibitem[{int()}]{integrals}
\bibinfo{note}{The prefactors in Eq. 2 are calculated by integrating the
  modeshapes and their derivatives as follows: $\int_0^1 \xi(x) (\xi(x)'
  (\xi(x)' \xi(x)'')')' \mathrm{d}x = 40.44$, $\int_0^1 \xi(x) \mathrm{d}x =
  0.78$ and $\int_0^1 \xi(x)(\xi'(x) \int_0^x \int_1^{x_1} \xi'(x{_2})^2
  \mathrm{d}x_2 \mathrm{d}x{_1})' dx = 4.60$}.
\bibitem[{pie()}]{piezo}
\bibinfo{note}{A different piezo actuator is used in this experiment.}
\end{thebibliography}

\newpage

\end{document}